\documentclass{article}
\usepackage{graphicx}
\usepackage{amsmath}
\usepackage{hyperref}
\usepackage{cite}
\usepackage{caption}
\usepackage{tabularx}
\usepackage{array}
\newcolumntype{Z}{>{\raggedright\arraybackslash}X}
\newcolumntype{Y}{>{\ttfamily\small\raggedright\arraybackslash}X}
\usepackage{placeins}

\usepackage{booktabs}
\usepackage{siunitx}

\title{Adaptive and Robust Cost-Aware Proof of Quality for Decentralized LLM Inference Networks}

\author{
  Arther Tian\textsuperscript{a},
  Alex Ding\textsuperscript{a,*},
  Frank Chen\textsuperscript{a}\\
  Simon Wu\textsuperscript{a},
  Aaron Chan\textsuperscript{a}\\
  \textsuperscript{a}DGrid AI\\[0.5em]
  \textsuperscript{*}Corresponding author: \texttt{alex.ding@dgrid.ai}
}

\date{}
\usepackage[margin=1in]{geometry}
\begin{document}
\maketitle

\begin{abstract}
\noindent
Decentralized large language model inference networks require lightweight mechanisms to reward high quality outputs under heterogeneous latency and cost. Proof of Quality provides scalable verification by sampling evaluator nodes that score candidate outputs, then aggregating their scores into a consensus signal that determines rewards. However, evaluator heterogeneity and malicious score manipulation can distort consensus and inflate payouts, which weakens incentive alignment in open participation settings. 

This paper extends a cost-aware Proof of Quality mechanism by adding adversary-resilient consensus formation. We study robust aggregation rules, including median and trimmed mean, and an adaptive trust-weighted consensus that updates evaluator weights from deviation signals. Using question answering and summarization workloads with a ground truth proxy for offline analysis, we quantify evaluator reliability and show strong variance across evaluators, including task-dependent misalignment that can invert correlations. We then evaluate robustness under four adversarial strategies, including noise injection, boosting, sabotage, and intermittent manipulation, across a sweep of malicious ratios and evaluator sample sizes. Our results show that robust aggregation improves consensus alignment with the ground truth proxy and reduces sensitivity to noisy and strategic attacks compared with simple averaging. We further characterize the operational trade-off introduced by evaluator sampling, where larger evaluator sets reduce evaluator rewards and increase payoff variance while inference rewards remain relatively stable in our configuration. These findings motivate robust consensus as a default component for cost-aware Proof of Quality and provide practical guidance for selecting evaluator sampling parameters under adversarial risk and resource constraints.
\end{abstract}

\section{Introduction}

Decentralized large language model inference aims to provide open access to advanced AI capabilities with transparency and censorship resistance, while avoiding single points of failure in centralized deployments. Achieving this goal requires a trustless mechanism to assess output quality at scale. Cryptographic verification for neural inference remains expensive for modern transformer models at interactive latency, so many practical systems rely on learned evaluation signals and aggregation to approximate quality at low cost. Proof of Quality, abbreviated as PoQ, follows this direction by collecting scores from lightweight evaluators and forming a network level judgment \cite{zhang2024poq}.

Our prior work introduced a cost-aware PoQ mechanism for decentralized LLM inference networks \cite{tian2025costawarepoq}. The key idea is to reward inference nodes and evaluator nodes using a joint objective that balances output quality and computational cost. We profiled efficiency across model choices and showed that evaluator selection can strongly affect alignment with task level quality. In particular, semantic textual similarity style bi-encoders can provide a strong signal with low overhead, which is consistent with the efficiency of sentence embedding architectures such as Sentence-BERT \cite{reimers2019sentence}.

This extended paper addresses a limitation that becomes unavoidable in open and permissionless deployments. A decentralized network must tolerate unreliable or adversarial evaluators. Even if inference nodes behave honestly, a fraction of evaluator nodes can manipulate scores to bias consensus. Such manipulations can boost preferred outputs, sabotage competitors, or inject noise that destabilizes rewards. Under these conditions, a consensus rule based on a simple mean is fragile, and the reward mechanism can be exploited to cause overpayment or to distort selection toward low quality outputs. These risks connect PoQ to classical Byzantine settings where a subset of participants deviates arbitrarily, and robust aggregation becomes central to safety and incentive alignment \cite{lamport1982byzantine,guerraoui2024byzantine}.

We propose an adversary-resilient extension of cost-aware PoQ that strengthens the evaluator layer while preserving the original incentive principle. The extension has two components. First, we introduce robust aggregation rules for consensus formation, including median and trimmed mean, which reduce sensitivity to extreme scores and have demonstrated value in adversarial learning settings \cite{pmlr-v80-yin18a,blanchard2017krum}. Second, we introduce an adaptive trust weighting mechanism that updates evaluator weights using observed deviation from consensus over time. Evaluators that repeatedly diverge receive reduced influence, while consistent evaluators retain or gain influence. This design aligns with a long line of work on combining judgments from sources of unknown expertise, where reliability is inferred from observed agreement patterns \cite{dawid1979em,raykar2010learningfromcrowds}.

\begin{figure}[t]
\centering
\includegraphics[width=0.8\columnwidth]{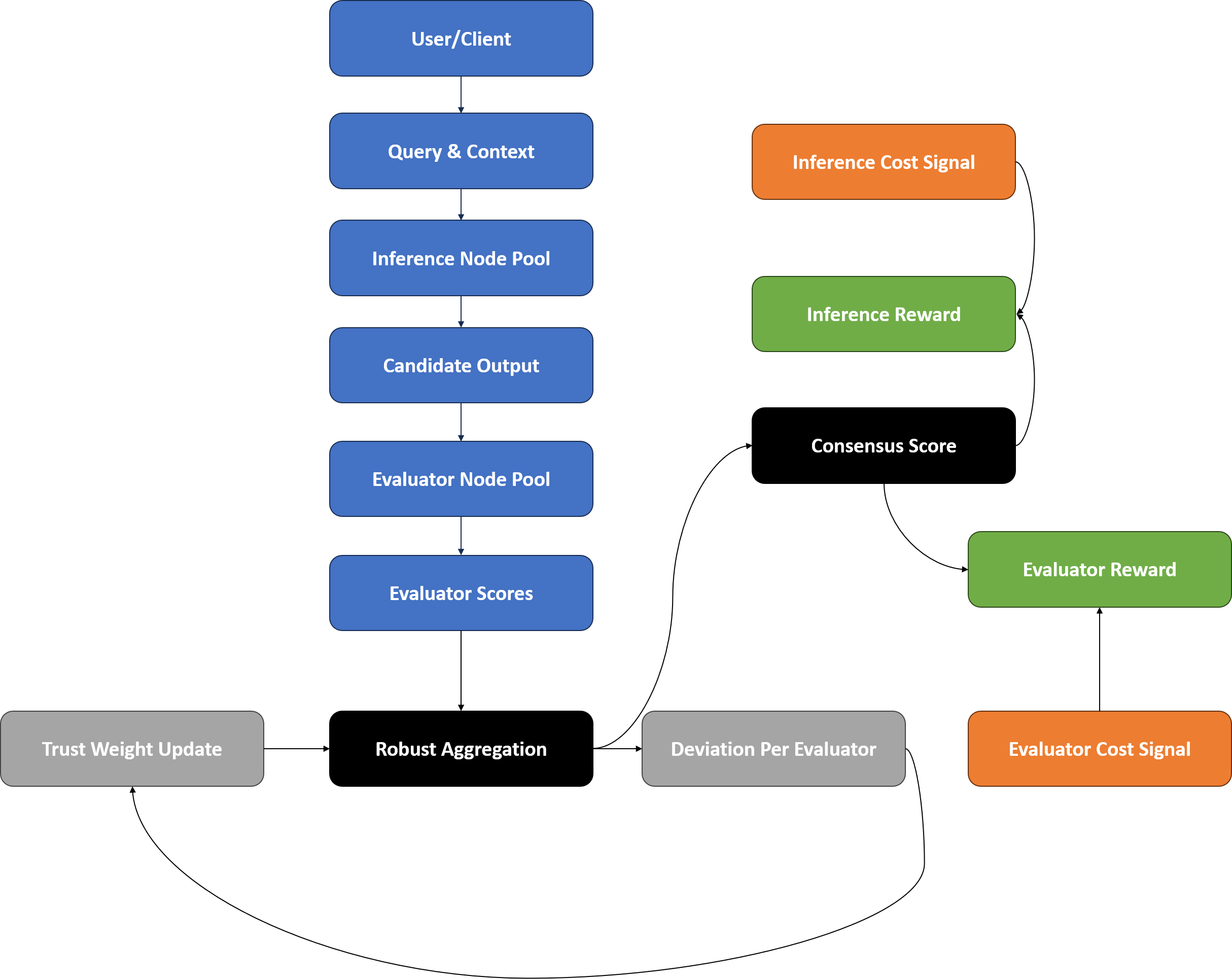}
\caption{System overview of adversary-resilient cost-aware PoQ. The figure highlights the main workflow from inference to evaluation, together with cost signals, reward assignment, robust aggregation, deviation signals, and trust weight updates.}
\label{fig:system_overview}
\end{figure}

\begin{figure}[t]
\centering
\includegraphics[width=\columnwidth]{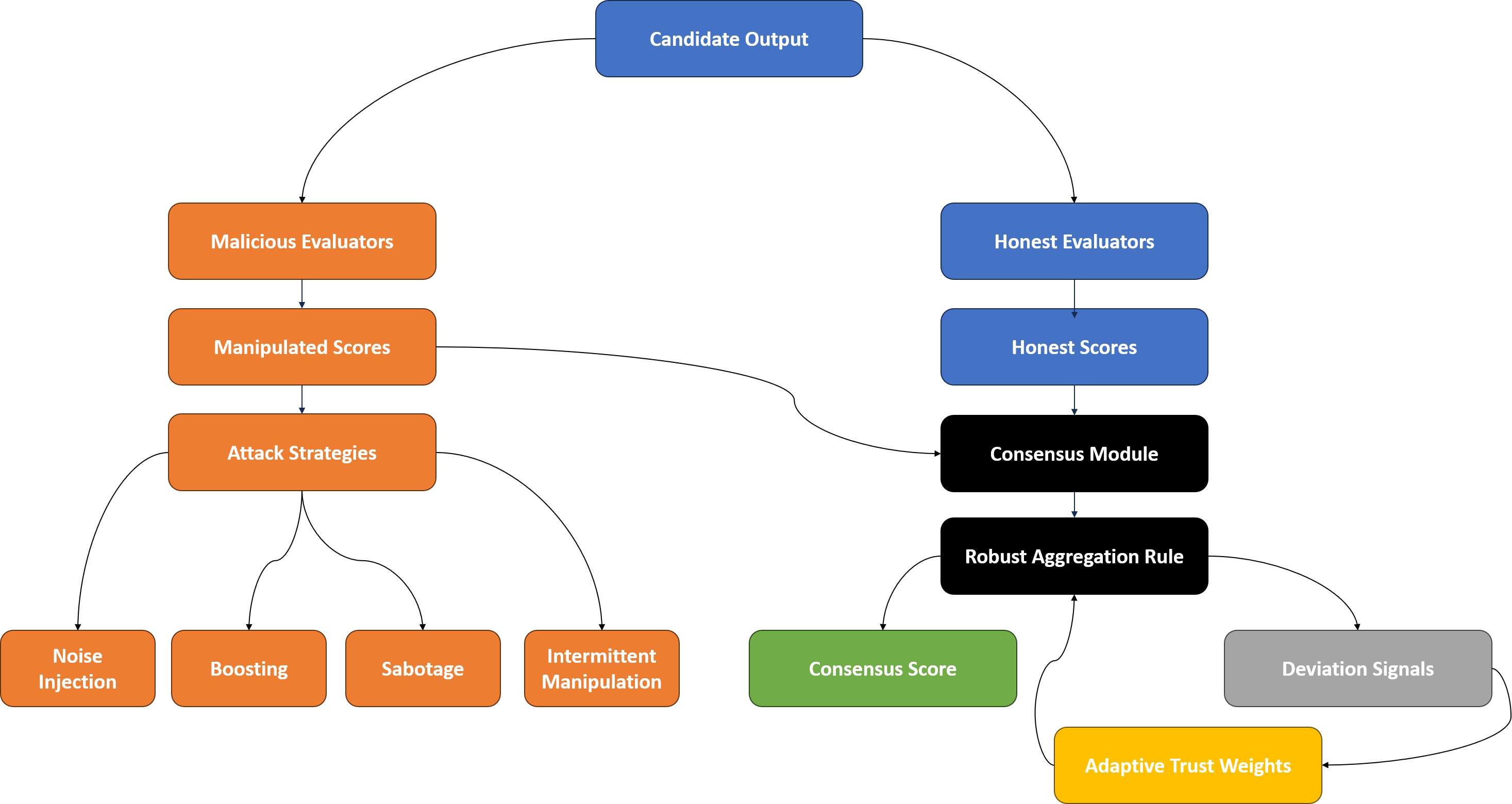}
\caption{Threat model for malicious evaluators. Malicious evaluators manipulate submitted scores via noise injection, boosting, sabotage, and intermittent strategies. The consensus module uses robust aggregation and deviation signals to support adaptive trust weights.}
\label{fig:threat_model}
\end{figure}

Figure~\ref{fig:system_overview} summarizes the end-to-end workflow and accounting signals, including cost normalization, consensus formation, and reward assignment for inference and evaluator nodes. Figure~\ref{fig:threat_model} outlines the malicious evaluator threat model and highlights where robust aggregation and deviation-driven trust weights mitigate score manipulation.

We evaluate the proposed mechanism using a Monte Carlo simulation engine driven by real evaluator score records and latency based cost normalization. The evaluator pool spans bi-encoder and cross-encoder architectures, and the inference pool includes multiple open source LLMs with distinct latency profiles. We quantify evaluator reliability by correlation against task level ground truth and compare aggregation rules by their alignment with ground truth. We then stress test robustness under multiple adversarial attack types and malicious ratios, and we analyze sensitivity to the number of evaluators sampled per job, denoted as $K$, which affects robustness and evaluation overhead.

This paper makes the following contributions.

\begin{itemize}
\item We extend cost-aware PoQ toward adversary resilience by integrating robust consensus rules and adaptive evaluator trust weighting, while preserving explicit cost awareness in rewards \cite{tian2025costawarepoq}.
\item We provide a reliability study of common lightweight evaluators across tasks, and we show how robust aggregation improves consensus alignment with ground truth under evaluator heterogeneity \cite{pmlr-v80-yin18a,dawid1979em}.
\item We introduce and evaluate a malicious evaluator threat model with multiple attack strategies and malicious ratios, reporting stability trends and failure modes across defenses \cite{fang2020localpoison,pmlr-v108-bagdasaryan20a}.
\item We present a sensitivity analysis over evaluator sampling size $K$, clarifying the trade-off between robustness, reward variance, and evaluation overhead for practical deployments.
\end{itemize}

\section{Related Work}

\subsection{Proof of Quality and Decentralized Inference}
Proof of Quality, abbreviated as PoQ, uses lightweight evaluators to score candidate outputs and aggregates these scores into a network level quality signal, enabling practical verification when cryptographic proof systems are costly at interactive latency \cite{zhang2024poq}. Our prior work extended PoQ with explicit cost awareness by incorporating latency-based cost signals into reward computation for inference nodes and evaluator nodes \cite{tian2025costawarepoq}. This paper further extends that line by focusing on adversarial evaluator robustness, which becomes critical in open participation settings where evaluators can be unreliable or malicious.

Decentralized learning and coordination have a long history, including federated learning as a common paradigm where multiple participants contribute updates under heterogeneous resources and incentives \cite{kairouz2021advances}. While our setting differs because evaluators do not train a shared model, many of the same concerns arise, including trust, participant heterogeneity, and susceptibility to manipulation. Secure aggregation and related techniques can reduce information leakage in collaborative settings, yet they do not by themselves address the correctness of submitted scores \cite{bonawitz2017practical}. In PoQ-style designs, robustness must primarily be achieved through aggregation rules and incentive mechanisms.

\subsection{Robust Aggregation and Byzantine-Resilient Learning}
Byzantine fault tolerance formalizes systems where a subset of participants can deviate arbitrarily \cite{lamport1982byzantine,castro1999practical}. In machine learning, Byzantine-resilient methods study how to aggregate potentially corrupted information, often gradients or model updates, to maintain learning progress \cite{blanchard2017krum,pmlr-v80-yin18a}. These works highlight that naive averaging can be brittle when adversaries control even a small fraction of participants, which motivates robust aggregation rules and defenses \cite{pmlr-v80-mhamdi18a,guerraoui2024byzantine}.

Robust statistics provides classical tools for limiting the influence of outliers, including Huber-style estimators and heavy-tail mean estimation \cite{huber1964robust,catoni2012challenging,devroye2016subgaussian}. Median and trimmed mean are widely used robust aggregators, and related results characterize their concentration and robustness in adversarial or heavy-tailed regimes \cite{minscher2015geometricmedian,lugosi2019heavytailedsurvey}. In our work, these ideas translate into robust score aggregation for PoQ, where malicious evaluators play a role analogous to Byzantine workers, and robust consensus reduces sensitivity to extreme score manipulations.

\subsection{Learning from Multiple Evaluators and Reliability Estimation}
Aggregating judgments from multiple sources with unknown expertise has been studied extensively in crowdsourcing and label aggregation. The Dawid--Skene model infers annotator error rates via maximum likelihood, enabling improved consensus labels under heterogeneous annotator quality \cite{dawid1979em}. Subsequent work extends this approach to supervised learning with noisy labels, jointly modeling worker reliability and classifier parameters \cite{raykar2010learningfromcrowds}. Other directions optimize label integration by estimating expertise and task difficulty, which can guide reweighting of annotators and improve label quality \cite{whitehill2009whosevote,zhou2012minimaxentropy}.

These formulations motivate the reliability component of our approach. In PoQ, evaluator nodes play the role of annotators, and their scores may be systematically biased or adversarial. Deviation-based trust updates provide a lightweight alternative to full probabilistic modeling, while still capturing the key idea that evaluators should be weighted by inferred reliability.

\subsection{Adversarial Manipulation of Evaluation and Reward Signals}
Adversarial behavior in decentralized settings often targets the aggregation layer. In federated learning, model poisoning and backdoor attacks show that a small number of malicious participants can significantly degrade or compromise the global model \cite{pmlr-v108-bagdasaryan20a,fang2020localpoison}. Sybil settings further amplify the threat by allowing an attacker to control multiple identities, which can undermine defenses that assume independent participants \cite{fung2020limitations}. Related analyses emphasize that adversaries can exploit aggregation rules, selection policies, and trust mechanisms when system assumptions are violated \cite{pmlr-v97-bhagoji19a}.

Our malicious evaluator model adapts these adversarial perspectives to PoQ. Instead of poisoning gradients, adversaries manipulate evaluation scores to bias consensus and rewards. This motivates evaluating robust aggregation and trust updates under multiple attack strategies and malicious ratios.

\subsection{Incentives, Reputation, and Peer Prediction}
Reputation systems provide a standard tool for encouraging honest behavior in open networks by aggregating historical interactions into trust scores, with EigenTrust as a well-known example in peer-to-peer systems \cite{kamvar2003eigentrust}. Mechanism design for truthful reporting offers complementary approaches. Peer prediction mechanisms elicit informative feedback without ground truth by rewarding reports that are predictively consistent with others \cite{miller2005peerprediction}. Bayesian truth serum provides another approach for truthful elicitation of subjective data \cite{prelec2004bayesian}.

PoQ shares the need to reward participants based on signals that can be noisy or strategic. Our cost-aware reward design follows the PoQ incentive structure \cite{tian2025costawarepoq}, while our robustness extensions focus on limiting the impact of strategic evaluators through robust aggregation and trust weighting, which are compatible with broader reputation and incentive ideas.

\subsection{LLMs as Judges and Judgment Bias}
LLM-based judging has become common for evaluating natural language generation, yet reliability concerns remain, including bias, task dependence, and susceptibility to prompting effects. Recent studies compare human and LLM judgments and characterize bias patterns that can affect evaluation outcomes \cite{chen-etal-2024-humans}. Other work analyzes factuality evaluation capabilities and reliability limits of LLM-based judges \cite{fu-etal-2023-large,wang-etal-2024-assessing}. In our pipeline, GPT-based judging is used as an auxiliary signal for analysis rather than a sole arbiter, and our threat model centers on adversarial evaluator nodes that directly manipulate the PoQ scoring layer.

\section{System Model and Threat Model}
\label{sec:system_threat}

\subsection{Network Roles and Workflow}
We consider a decentralized inference network with two types of participants. Inference nodes host an LLM and generate candidate outputs. Evaluator nodes host lightweight scoring models and produce quality scores for candidate outputs. A job $t$ consists of an input query and context $x_t$, a selected inference model $f_t$ from an inference pool $\mathcal{F}$, and a generated output $y_t$.

For each job, the protocol samples a subset of evaluators $\mathcal{E}_t \subseteq \mathcal{E}$ with size $K$. Each evaluator $e \in \mathcal{E}_t$ returns a scalar score $s_{t,e} \in [0,10]$ for output $y_t$. The network aggregates these scores into a consensus score $c_t \in [0,10]$ using a defense rule. The consensus score is then used to compute rewards for the selected inference node and the participating evaluators.

The protocol maintains a nonnegative trust weight $w_e$ for each evaluator. For consensus computation, we use a normalized weight vector so that the average weight is $1$ across the full evaluator pool $\mathcal{E}$,
\begin{equation}
\tilde{w}_e \;=\; w_e \cdot \frac{|\mathcal{E}|}{\sum_{j \in \mathcal{E}} w_j}.
\label{eq:normalized_weight}
\end{equation}
This normalization improves stability and makes weights comparable across runs with different absolute scales.

We also define a normalized quality score $q_t \in [0,1]$ for reward computation,
\begin{equation}
q_t \;=\; \frac{c_t}{10}.
\label{eq:quality_norm}
\end{equation}

\subsection{Cost Model and Normalization}
Each inference model $f \in \mathcal{F}$ and each evaluator model $e \in \mathcal{E}$ has an empirical average latency measured on a shared benchmark workload. Let $L_f$ denote the average latency of inference model $f$ and let $L_e$ denote the average latency of evaluator $e$. We convert latency into a unitless cost signal in $[0,1]$ using min max normalization within each role,
\begin{equation}
C_f \;=\; \frac{L_f - \min_{f' \in \mathcal{F}} L_{f'}}{\max_{f' \in \mathcal{F}} L_{f'} - \min_{f' \in \mathcal{F}} L_{f'}},
\label{eq:inference_cost_norm}
\end{equation}
\begin{equation}
C_e \;=\; \frac{L_e - \min_{e' \in \mathcal{E}} L_{e'}}{\max_{e' \in \mathcal{E}} L_{e'} - \min_{e' \in \mathcal{E}} L_{e'}}.
\label{eq:evaluator_cost_norm}
\end{equation}
These normalized costs are used directly in the reward functions to encourage higher quality per unit cost.

\subsection{Consensus Formation}
Given a set of scores $\{s_{t,e}\}_{e \in \mathcal{E}_t}$, the consensus module produces $c_t$. We evaluate four consensus rules.

The simple mean uses unweighted averaging,
\begin{equation}
c_t^{\mathrm{mean}} \;=\; \frac{1}{K} \sum_{e \in \mathcal{E}_t} s_{t,e}.
\label{eq:consensus_mean}
\end{equation}

The median uses the sample median over the $K$ scores,
\begin{equation}
c_t^{\mathrm{med}} \;=\; \mathrm{median}\Bigl(\{s_{t,e}\}_{e \in \mathcal{E}_t}\Bigr).
\label{eq:consensus_median}
\end{equation}

The trimmed mean removes a fraction of extreme scores at both ends and averages the remaining values. Let the sorted scores be $s_{t,(1)} \le \dots \le s_{t,(K)}$ and let $m = \max\{1, \lfloor \gamma K \rfloor\}$ for a trim ratio $\gamma \in (0,0.5)$. The trimmed mean is
\begin{equation}
c_t^{\mathrm{trim}} \;=\; \frac{1}{K - 2m}\sum_{i=m+1}^{K-m} s_{t,(i)}.
\label{eq:consensus_trimmed}
\end{equation}

The adaptive weighted mean uses trust weights to reduce the influence of unreliable evaluators,
\begin{equation}
c_t^{\mathrm{w}} \;=\; \frac{\sum_{e \in \mathcal{E}_t} \tilde{w}_e \, s_{t,e}}{\sum_{e \in \mathcal{E}_t} \tilde{w}_e}.
\label{eq:consensus_weighted}
\end{equation}
In our implementation, weights are updated online using deviation signals described next.

\subsection{Reward Design for Inference and Evaluator Nodes}
Inference rewards combine normalized quality and normalized cost. Let $\alpha_f$ and $\beta_f$ be reward weights and let $\tau$ be a quality threshold. The base inference reward is
\begin{equation}
R^{\mathrm{base}}_{t,f} \;=\; \alpha_f q_t - \beta_f C_{f_t}.
\label{eq:inference_reward_base}
\end{equation}
We apply a threshold penalty when quality falls below $\tau$,
\begin{equation}
R^{\mathrm{pen}}_{t,f} \;=\; 
\begin{cases}
\bigl(\tau - q_t\bigr)^2, & q_t < \tau,\\
0, & q_t \ge \tau,
\end{cases}
\label{eq:inference_penalty}
\end{equation}
and we add an efficiency bonus that favors higher quality at lower cost. Let $\eta$ be a bonus weight and let $B_{\max}$ cap the bonus,
\begin{equation}
R^{\mathrm{bon}}_{t,f} \;=\; \min\Bigl(\eta \, q_t \, (1 - C_{f_t}),\, B_{\max}\Bigr).
\label{eq:inference_bonus}
\end{equation}
The final inference reward is
\begin{equation}
R_{t,f} \;=\; R^{\mathrm{base}}_{t,f} \;-\; R^{\mathrm{pen}}_{t,f} \;+\; R^{\mathrm{bon}}_{t,f}.
\label{eq:inference_reward_final}
\end{equation}

Evaluator rewards depend on agreement with consensus and evaluation cost. We define a deviation and a closeness score,
\begin{equation}
d_{t,e} \;=\; \frac{|s_{t,e} - c_t|}{10},
\label{eq:deviation}
\end{equation}
\begin{equation}
a_{t,e} \;=\; \max\{0,\, 1 - d_{t,e}\}.
\label{eq:closeness}
\end{equation}
Let $\alpha_m$ and $\beta_m$ be evaluator reward weights. The evaluator reward is
\begin{equation}
R_{t,e} \;=\; \alpha_m a_{t,e} \;-\; \beta_m C_e.
\label{eq:evaluator_reward}
\end{equation}

\subsection{Adaptive Trust Weight Updates}
The protocol updates evaluator trust weights using deviation signals. Let $\lambda$ be a learning rate. The multiplicative update is
\begin{equation}
w_e \leftarrow \mathrm{clip}\Bigl(w_e \cdot \bigl(1 + \lambda (0.5 - d_{t,e})\bigr),\, w_{\min},\, w_{\max}\Bigr),
\label{eq:weight_update}
\end{equation}
where $\mathrm{clip}$ clamps the updated weight into the interval $[w_{\min}, w_{\max}]$. This rule increases trust when the deviation is smaller than $0.5$ and decreases trust when the deviation is larger than $0.5$, while preventing unbounded growth.

\subsection{Adversarial Evaluator Capabilities}
We consider an adversary that controls a subset of evaluators $\mathcal{M} \subseteq \mathcal{E}$ with malicious ratio $\rho = |\mathcal{M}|/|\mathcal{E}|$. The adversary can submit manipulated scores in place of honest scores for any job where a malicious evaluator is sampled. The adversary does not modify inference outputs, measured costs, or the consensus rule, and can be treated as fully informed about protocol parameters.

We evaluate four attack strategies that model common manipulations of scalar score signals.

Random noise injection perturbs the honest score by bounded noise,
\begin{equation}
s'_{t,e} \;=\; \mathrm{clip}\bigl(s_{t,e} + u_{t,e},\, 0,\, 10\bigr),
\qquad
u_{t,e} \sim \mathrm{Unif}[-r, r].
\label{eq:attack_noise}
\end{equation}

Boosting increases the score by a fixed bias $b$,
\begin{equation}
s'_{t,e} \;=\; \min\{10,\, s_{t,e} + b\}.
\label{eq:attack_boost}
\end{equation}

Sabotage decreases the score by the same type of bias,
\begin{equation}
s'_{t,e} \;=\; \max\{0,\, s_{t,e} - b\}.
\label{eq:attack_sabotage}
\end{equation}

Strategic manipulation applies a larger deviation $\Delta$ only with probability $p$, with random direction,
\begin{equation}
s'_{t,e} \;=\;
\begin{cases}
\mathrm{clip}\bigl(s_{t,e} + \xi_{t,e}\Delta,\, 0,\, 10\bigr), & \text{with probability } p,\\
s_{t,e}, & \text{with probability } 1-p,
\end{cases}
\qquad
\xi_{t,e} \in \{-1, +1\}.
\label{eq:attack_strategic}
\end{equation}

In all cases, the protocol uses the submitted score $s'_{t,e}$ in consensus and rewards. Our evaluation varies $\rho$, $K$, and the consensus rule to quantify robustness under these adversarial behaviors.

\section{Adversary-Resilient Cost-Aware PoQ Mechanism}
\label{sec:robust_poq}

This section presents the protocol components that extend cost-aware PoQ toward adversary resilience. The core idea is to preserve the original cost-aware incentive structure while hardening the consensus layer against score manipulation. The mechanism combines robust aggregation rules with an adaptive trust weighting scheme that downweights evaluators that frequently deviate from consensus.

\subsection{Robust Consensus Rules}
\label{subsec:robust_consensus}

Given a job $t$ with evaluator subset $\mathcal{E}_t$ of size $K$ and reported scores $\{s_{t,e}\}_{e \in \mathcal{E}_t}$, the consensus module outputs a consensus score $c_t$. We consider four consensus rules, all implemented as drop-in alternatives in the same scoring interface.

Simple mean uses Eq.~\ref{eq:consensus_mean} and is optimal when evaluator scores are unbiased and approximately independent. However, it is sensitive to extreme values and can be exploited by boosting or sabotage style attacks.

Median uses Eq.~\ref{eq:consensus_median} and reduces the influence of large outliers. It is a natural choice when a portion of evaluators may be unreliable.

Trimmed mean uses Eq.~\ref{eq:consensus_trimmed} and removes a fraction of extreme scores on both ends before averaging. In our implementation the trim ratio $\gamma$ is a fixed hyperparameter. Trimmed mean often preserves more information than the median while still limiting adversarial impact, which is a common robustness goal in Byzantine settings \cite{pmlr-v80-yin18a,blanchard2017krum}.

Adaptive weighted mean uses Eq.~\ref{eq:consensus_weighted} with normalized trust weights from Eq.~\ref{eq:normalized_weight}. This rule generalizes simple mean by allowing the protocol to reduce the influence of evaluators that appear unreliable over time.

\subsection{Deviation Signals and Adaptive Trust Weights}
\label{subsec:trust_weights}

Robust aggregation mitigates single-round manipulation, yet persistent adversaries can still bias consensus if they participate frequently. To address this, the protocol maintains an online trust weight $w_e$ for each evaluator $e \in \mathcal{E}$. Trust weights are updated using deviation signals computed at each round.

For each evaluator $e \in \mathcal{E}_t$, we compute the normalized deviation using Eq.~\ref{eq:deviation}. The deviation is recorded for analysis and also drives a multiplicative trust update,
\begin{equation}
w_e \leftarrow \mathrm{clip}\Bigl(w_e \cdot \bigl(1 + \lambda (0.5 - d_{t,e})\bigr),\, w_{\min},\, w_{\max}\Bigr),
\label{eq:trust_update}
\end{equation}
where $\lambda > 0$ is a learning rate and $[w_{\min}, w_{\max}]$ defines the allowed trust range. This update increases weight when the evaluator is closer to consensus and decreases weight when the evaluator deviates. The midpoint $0.5$ ensures that the update is symmetric in the sense that deviations below $0.5$ increase trust and deviations above $0.5$ decrease trust.

The use of deviation-driven trust weights is motivated by reliability estimation in multi-annotator settings, where the goal is to infer which sources are consistent with the collective signal \cite{dawid1979em,raykar2010learningfromcrowds}. In PoQ, this approach has two practical advantages. It is lightweight and requires no additional supervision. It also directly targets the attack surface, since manipulation changes deviation patterns even when the adversary adapts across jobs.

\subsection{Cost-Aware Rewards with Robust Consensus}
\label{subsec:rewards}

The protocol computes a normalized quality score $q_t$ from consensus using Eq.~\ref{eq:quality_norm} and then applies cost-aware reward functions for inference nodes and evaluators.

For inference nodes, the reward uses the same structure as the cost-aware baseline, including a quality threshold penalty and a bounded efficiency bonus. The final inference reward follows Eq.~\ref{eq:inference_reward_final}, where the base term, penalty, and bonus are defined in Eq.~\ref{eq:inference_reward_base}, Eq.~\ref{eq:inference_penalty}, and Eq.~\ref{eq:inference_bonus}. This design prevents reward inflation for extremely low cost models by avoiding division-based efficiency terms and by capping the bonus.

For evaluators, rewards depend on agreement with consensus and evaluation cost. We compute closeness using Eq.~\ref{eq:closeness} and compute evaluator reward using Eq.~\ref{eq:evaluator_reward}. This encourages evaluators to provide scores that are consistent with the collective signal while remaining efficient. Under attack, this reward also discourages persistent manipulation because adversarial scores increase deviation and reduce closeness.

\subsection{Protocol for One PoQ Round}
\label{subsec:protocol_round}

We now summarize the protocol for a single PoQ round, which is the unit used in our Monte Carlo simulations.

\begin{enumerate}
\item Sample a record that contains a model key and precomputed normalized evaluator scores.
\item Select an inference model key $f_t$ from the record and retrieve its normalized inference cost $C_{f_t}$ using Eq.~\ref{eq:inference_cost_norm}.
\item Identify evaluators that have available normalized scores for this record, then sample a subset $\mathcal{E}_t$ of size $K$.
\item For each evaluator $e \in \mathcal{E}_t$, obtain the score $s_{t,e}$. If $e$ is malicious, replace the honest score with a manipulated score using the threat model in Eq.~\ref{eq:attack_noise} through Eq.~\ref{eq:attack_strategic}.
\item Compute consensus score $c_t$ using one of the rules in Section~\ref{subsec:robust_consensus}. If the rule uses trust weights, provide normalized weights from Eq.~\ref{eq:normalized_weight}.
\item Compute the inference reward using Eq.~\ref{eq:inference_reward_final} and record it for $f_t$.
\item For each evaluator $e \in \mathcal{E}_t$, compute deviation using Eq.~\ref{eq:deviation}, compute closeness using Eq.~\ref{eq:closeness}, compute evaluator reward using Eq.~\ref{eq:evaluator_reward}, and record the reward.
\item Update evaluator trust weights using Eq.~\ref{eq:trust_update}.
\end{enumerate}

This round-level procedure is repeated for a fixed number of rounds $T$ to estimate expected rewards, reward variance, and robustness properties under different malicious ratios, attack types, and consensus defenses.

\subsection{Implementation Notes}
\label{subsec:implementation_notes}

The mechanism is implemented in a modular form. The consensus rule is selected by a method identifier and can be switched without changing the rest of the pipeline. The reward module encapsulates all reward parameters, including threshold and bonus caps. The state module maintains trust weights, deviation histories, and cumulative reward statistics, enabling both robustness evaluation and correlation analysis.

Two hyperparameters directly control robustness and overhead. The evaluator sample size $K$ increases the amount of scoring information per job but also increases evaluation cost. The trust learning rate $\lambda$ affects how quickly the system reacts to deviations, and it can trade off responsiveness against stability. We examine both factors empirically in Section~\ref{sec:results}.

\section{Experimental Setup}
\label{sec:exp_setup}

\subsection{Datasets and Tasks}
We evaluate on two standard natural language generation tasks that reflect common decentralized inference workloads. For extractive style question answering, we use the SQuAD v1.1 development set \cite{rajpurkar-etal-2016-squad}. For news summarization, we use CNN and Daily Mail articles paired with reference highlights \cite{hermann2015teaching}. We build a unified task file by sampling 200 examples per dataset with a fixed random seed, resulting in 400 unique inputs.

\subsection{Inference Model Pool and Generation Protocol}
We use five open source instruction tuned LLMs as inference nodes, covering a range of parameter scales and efficiency profiles. Each inference model generates an output for every sampled task instance, producing 2000 generation records in total. Prompts are task-specific and fixed across models. For question answering, the prompt requests a concise answer conditioned on the provided context and question. For summarization, the prompt requests a concise summary in 2 to 3 sentences. We run decoding deterministically with greedy generation and a fixed maximum generation length.

\subsection{Ground Truth Proxy and Record Construction}
Each generation record contains the model output and a reference answer or reference summary. To obtain a uniform ground truth proxy across tasks, we compute token-level F1 following the SQuAD style normalization procedure, then rescale to the range $[0,10]$ as
\begin{equation}
\mathrm{GT}_{t} \;=\; 10 \cdot \mathrm{F1}(y_t, y_t^{\star}),
\label{eq:gt_score}
\end{equation}
where $y_t$ is the model output and $y_t^{\star}$ is the reference. This score is used only for offline analysis, including correlation studies and robustness evaluation.

\subsection{Evaluator Pool and Score Normalization}
We implement a heterogeneous evaluator pool with two cross-encoder models and three bi-encoder models. Each evaluator produces a raw similarity score for the pair formed by the reference and the model output. Since raw score scales differ across evaluators, we apply min max normalization per evaluator across all 2000 records and map scores into $[0,10]$. This produces a comparable score field for each evaluator, which is the input to the consensus module and adversarial perturbations.

\subsection{Efficiency Profiling and Cost Normalization}
We measure average latency, throughput, and peak GPU memory for both inference models and evaluator models using a shared workload sampled from the task file. For inference models, we warm up the generation pipeline and then measure average per prompt latency with a fixed generation length. For evaluators, we warm up and then score the full batch of pairs to obtain average per pair latency. We transform latency into a unitless cost signal in $[0,1]$ using min max normalization within inference models and within evaluator models, consistent with Eq.~\ref{eq:inference_cost_norm} and Eq.~\ref{eq:evaluator_cost_norm}.

\subsection{Monte Carlo Simulation Protocol}
We evaluate the mechanism using the PoQ simulation engine. Each simulation round samples one generation record uniformly at random and selects an evaluator subset of size $K$ from the available evaluator pool. The engine computes a consensus score using the selected defense rule and then computes rewards for the inference node and participating evaluators. Trust weights are maintained in the engine state and updated online using deviation signals.

Unless stated otherwise, we run $T=5000$ rounds per configuration. The default evaluator sample size is $K=3$. The reward parameters follow the cost-aware baseline, including an efficiency bonus cap, and a quality threshold penalty. The trust weighting component uses multiplicative updates with an initial weight of $1.0$ and clamping bounds, with a learning rate set by the configuration.

\subsection{Adversarial Settings and Sensitivity Sweeps}
To model malicious evaluators, we randomly select a fraction $\rho$ of evaluators and replace their scores using the attack models in Section~\ref{sec:system_threat}. We evaluate four attack types, including random noise injection, boosting, sabotage, and strategic intermittent manipulation. For robustness comparison we test multiple malicious ratios and multiple defense rules, including simple mean, median, trimmed mean, and adaptive weighted consensus.

We also run two sensitivity sweeps. First, we vary $K$ from 1 to the full evaluator pool size to quantify the trade-off between robustness and evaluation overhead. Second, we vary the malicious ratio with finer granularity to identify stability regimes and failure modes as adversarial participation increases.

\section{Results}
\label{sec:results}

Unless stated otherwise, all results use $T=5000$ Monte Carlo rounds per configuration and $K=3$ evaluators sampled per job. We report averages over rounds, and we use the same cost normalization and reward parameters described in Section~\ref{sec:system_threat} and Section~\ref{sec:exp_setup}.

\subsection{Baseline Rewards and Efficiency Trends}
\label{subsec:results_baseline}

\begin{figure}[t]
\centering
\includegraphics[width=\columnwidth]{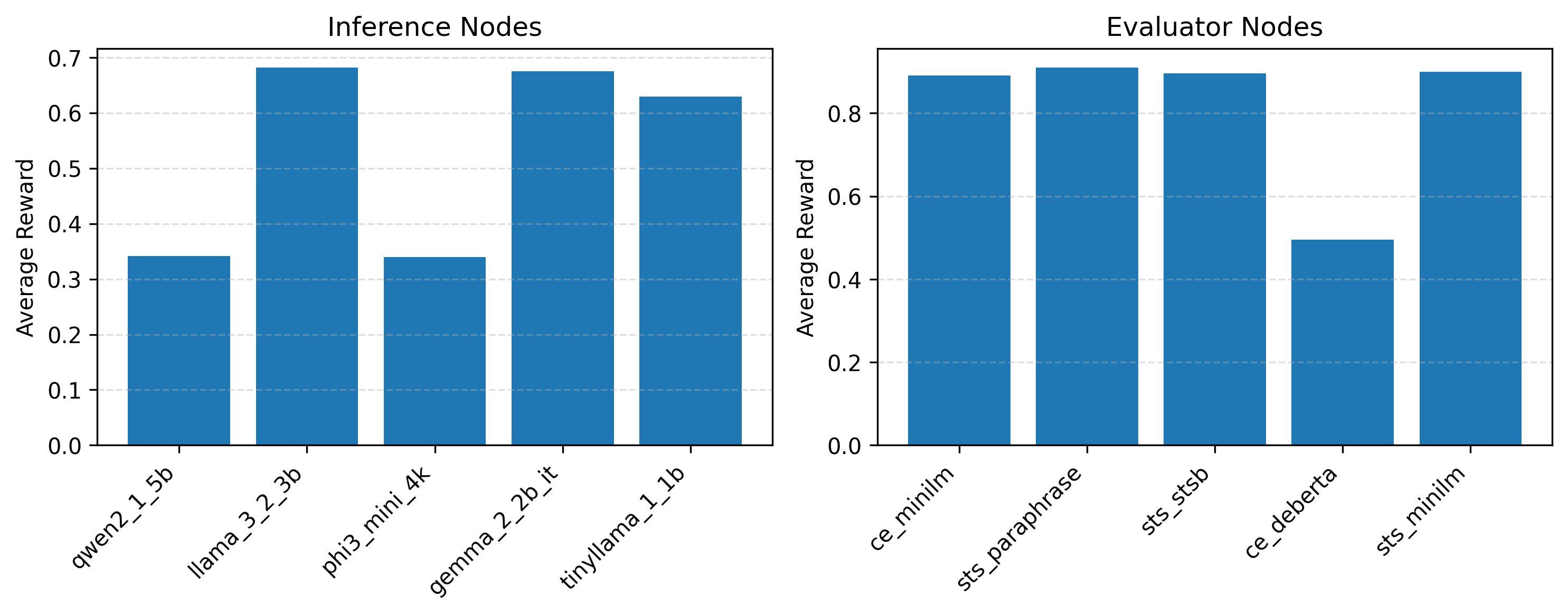}
\caption{Baseline rewards for inference nodes and evaluator nodes under cost-aware PoQ.}
\label{fig:baseline_rewards}
\end{figure}

We first report the baseline performance of cost-aware PoQ with heterogeneous evaluators and the default consensus rule. Figure~\ref{fig:baseline_rewards} visualizes baseline reward statistics across inference nodes and evaluators. Table~\ref{tab:baseline_inference} summarizes average inference rewards together with ground truth proxy quality and normalized cost. Table~\ref{tab:baseline_evaluator} summarizes evaluator rewards, average deviation from consensus, and normalized evaluation cost.

\begin{table}[t]
\centering
\small
\setlength{\tabcolsep}{5pt}
\caption{Baseline per inference model outcomes. Avg GT is the average normalized ground truth proxy in $[0,1]$. Cost is the normalized latency cost in $[0,1]$.}
\label{tab:baseline_inference}
\begin{tabular}{@{}l
S[table-format=1.3]
S[table-format=1.3]
S[table-format=1.3]
r@{}}
\toprule
Model & {Avg reward} & {Avg GT} & {Cost} & Jobs \\
\midrule
llama\_3\_2\_3b   & 0.682 & 0.541 & 0.223 & 1033 \\
gemma\_2\_2b\_it  & 0.675 & 0.519 & 0.194 & 1012 \\
tinyllama\_1\_1b  & 0.630 & 0.205 & 0.000 & 974  \\
qwen2\_1\_5b      & 0.341 & 0.165 & 1.000 & 1007 \\
phi3\_mini\_4k    & 0.340 & 0.129 & 0.843 & 974  \\
\bottomrule
\end{tabular}
\end{table}

\begin{table}[t]
\centering
\small
\setlength{\tabcolsep}{5pt}
\caption{Baseline evaluator outcomes. Avg deviation uses Eq.~\ref{eq:deviation}. Cost is the normalized latency cost in $[0,1]$.}
\label{tab:baseline_evaluator}
\begin{tabular}{@{}l
S[table-format=1.3]
S[table-format=1.3]
S[table-format=1.3]
r@{}}
\toprule
Evaluator & {Avg reward} & {Avg deviation} & {Cost} & Jobs \\
\midrule
sts\_paraphrase & 0.910 & 0.090 & 0.020 & 3037 \\
sts\_minilm     & 0.900 & 0.094 & 0.025 & 2969 \\
sts\_stsb       & 0.896 & 0.087 & 0.041 & 3006 \\
ce\_minilm      & 0.890 & 0.101 & 0.000 & 2983 \\
ce\_deberta     & 0.496 & 0.213 & 1.000 & 3005 \\
\bottomrule
\end{tabular}
\end{table}

Two patterns are consistent across runs. First, cost aware rewards rank inference models by a combination of quality and latency. Models with strong quality at moderate cost achieve the highest rewards, while high latency models are penalized even when they produce nontrivial quality. Second, evaluator rewards are strongly driven by agreement with the consensus score and by evaluation cost. The cross encoder evaluator with the highest normalized latency cost shows the lowest reward and the largest deviation, while the bi encoder evaluators provide high reward with low cost.

\subsection{Evaluator Reliability and Correlation Analysis}
\label{subsec:results_correlation}

We next quantify evaluator reliability by correlation against the ground truth proxy score. Table~\ref{tab:evaluator_correlation} reports Pearson and Spearman correlations for each evaluator over all records and within each task.

\begin{figure}[t]
\centering
\includegraphics[width=\columnwidth]{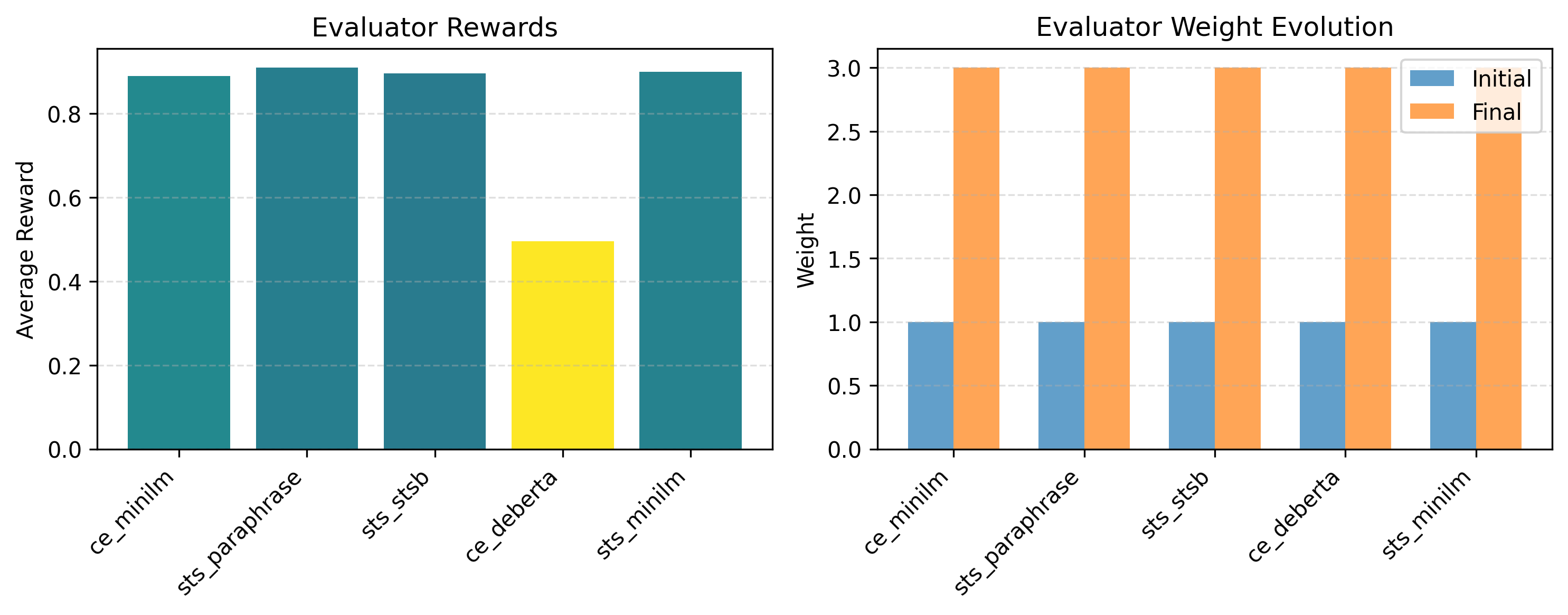}
\caption{Evaluator reliability analysis, including task level trends and deviation statistics.}
\label{fig:evaluator_analysis}
\end{figure}

\begin{figure}[t]
\centering
\includegraphics[width=0.8\columnwidth]{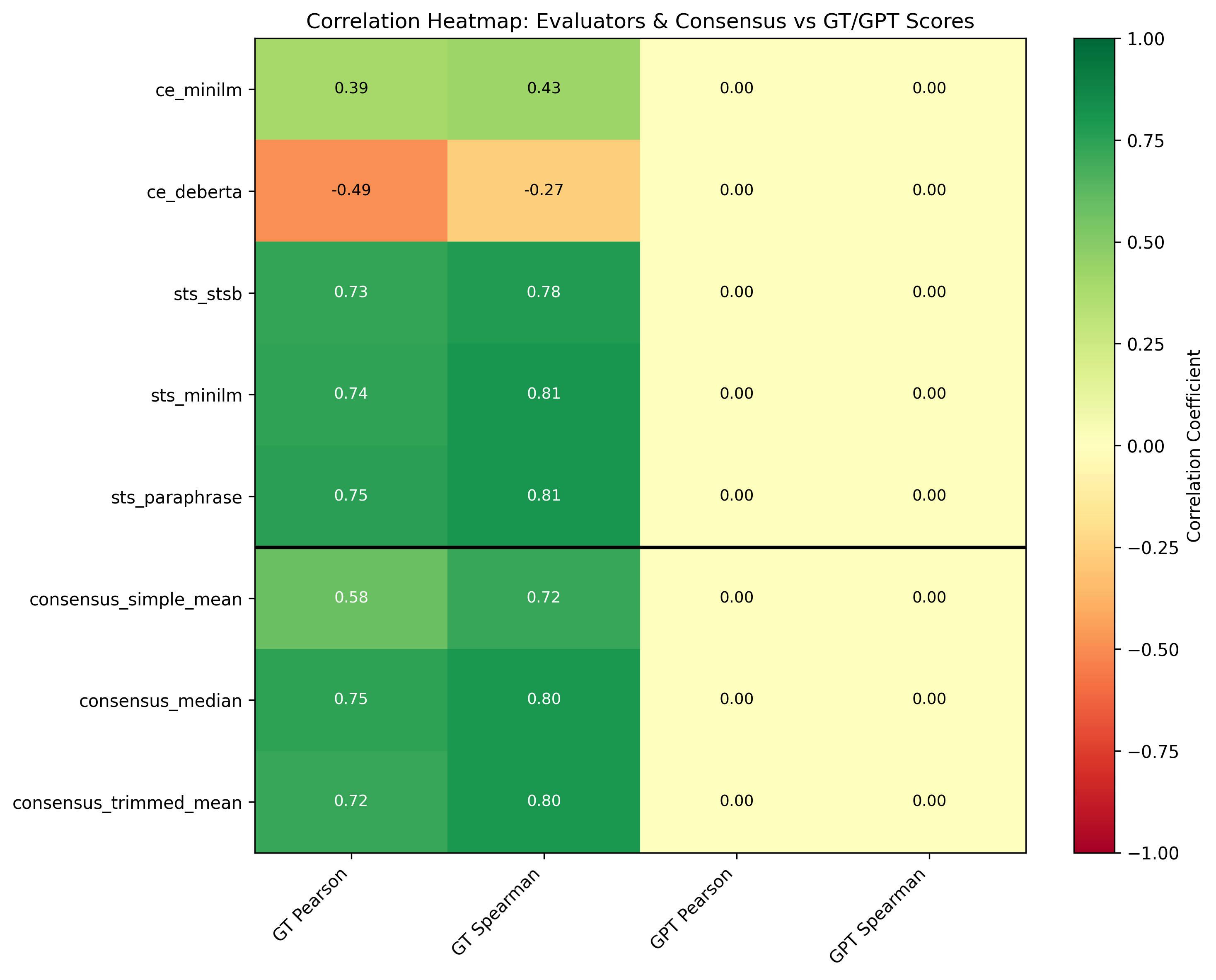}
\caption{Correlation heatmap between evaluator signals, consensus variants, and the ground truth proxy.}
\label{fig:correlation_heatmap}
\end{figure}

\begin{table}[t]
\centering
\caption{Correlation between evaluator scores and the ground truth proxy. Pearson and Spearman are computed on normalized scores in $[0,10]$. Task splits are QA and summarization.}
\label{tab:evaluator_correlation}
\begin{tabularx}{\columnwidth}{lrrrr}
\hline
Evaluator & Pearson all & Spearman all & Pearson QA & Pearson sum \\
\hline
ce\_minilm      & 0.392 & 0.426 & 0.422 & 0.425 \\
ce\_deberta     & -0.487 & -0.267 & -0.635 & 0.071 \\
sts\_stsb       & 0.733 & 0.776 & 0.863 & 0.433 \\
sts\_minilm     & 0.737 & 0.812 & 0.877 & 0.576 \\
sts\_paraphrase & 0.754 & 0.806 & 0.872 & 0.519 \\
\hline
\end{tabularx}
\end{table}

The bi encoder evaluators show strong correlation with the ground truth proxy, with the best overall Pearson correlation above 0.75. Correlations are task dependent, where QA correlations are consistently higher than summarization correlations. In contrast, one cross encoder evaluator exhibits negative correlation on QA, which indicates that evaluator selection can invert incentives if the scoring signal is misaligned with task quality. Figure~\ref{fig:evaluator_analysis} and Figure~\ref{fig:correlation_heatmap} provide additional views of these relationships.

\subsection{Consensus Alignment with Ground Truth}
\label{subsec:results_consensus}

We evaluate consensus variants by their correlation with ground truth. Table~\ref{tab:consensus_correlation} reports correlation results for simple mean, median, and trimmed mean computed over the full evaluator pool scores per record. Median improves Pearson correlation from 0.581 for simple mean to 0.748, and trimmed mean provides a comparable improvement.

\begin{figure}[t]
\centering
\includegraphics[width=\columnwidth]{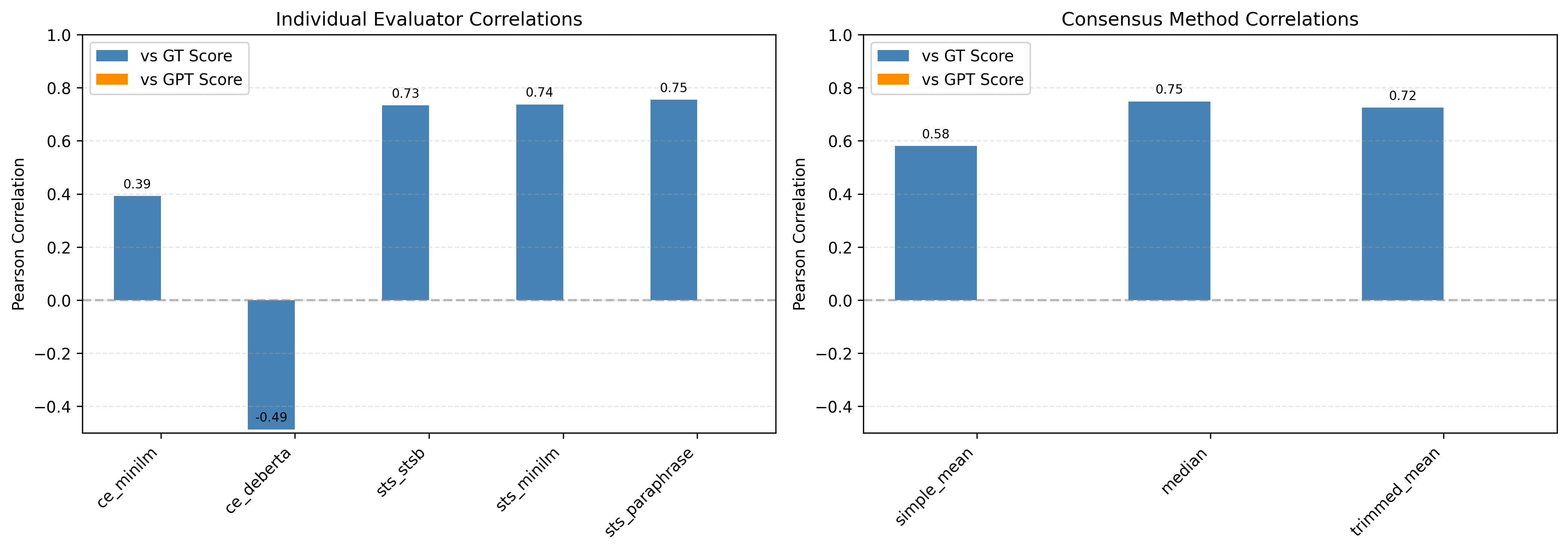}
\caption{Correlation of individual evaluators and consensus methods against the ground truth proxy.}
\label{fig:correlation_bar}
\end{figure}

\begin{table}[t]
\centering
\caption{Correlation between consensus scores and the ground truth proxy. Scores are aggregated over all available evaluator signals per record.}
\label{tab:consensus_correlation}
\begin{tabularx}{\columnwidth}{lrrrr}
\hline
Consensus & Pearson all & Spearman all & Pearson QA & Pearson sum \\
\hline
simple\_mean  & 0.581 & 0.720 & 0.742 & 0.514 \\
median        & 0.748 & 0.801 & 0.865 & 0.542 \\
trimmed\_mean & 0.725 & 0.798 & 0.844 & 0.545 \\
\hline
\end{tabularx}
\end{table}

These results motivate robust aggregation as a default defense in PoQ style systems, since it can recover alignment close to the best evaluator while reducing dependence on any single evaluator.

\subsection{Adversarial Robustness Across Attack Types}
\label{subsec:results_adversarial}

\begin{figure}[t]
\centering
\includegraphics[width=0.8\columnwidth]{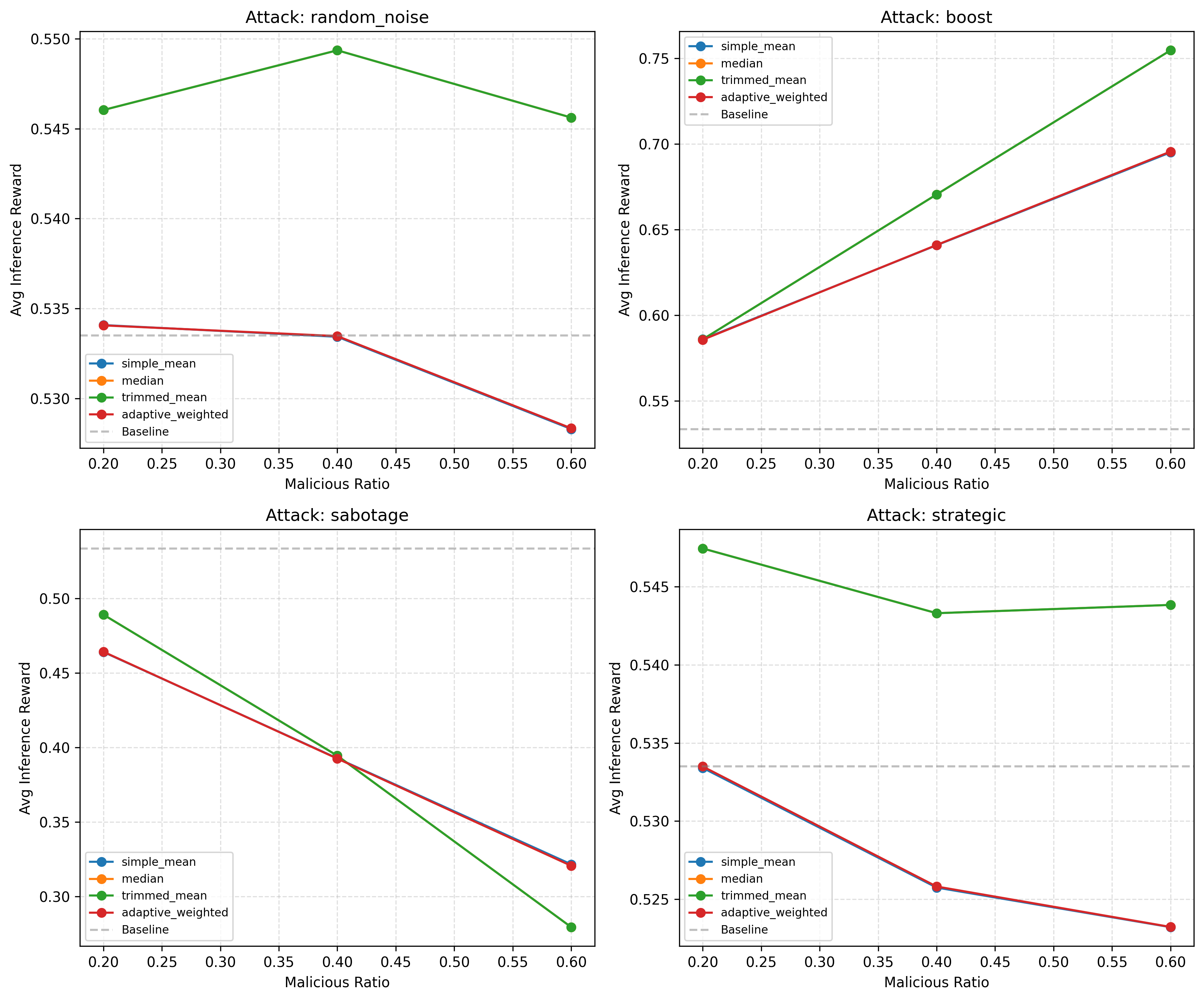}
\caption{Defense comparison under adversarial evaluators across attack types.}
\label{fig:adversarial_defense}
\end{figure}

\begin{figure}[t]
\centering
\includegraphics[width=0.8\columnwidth]{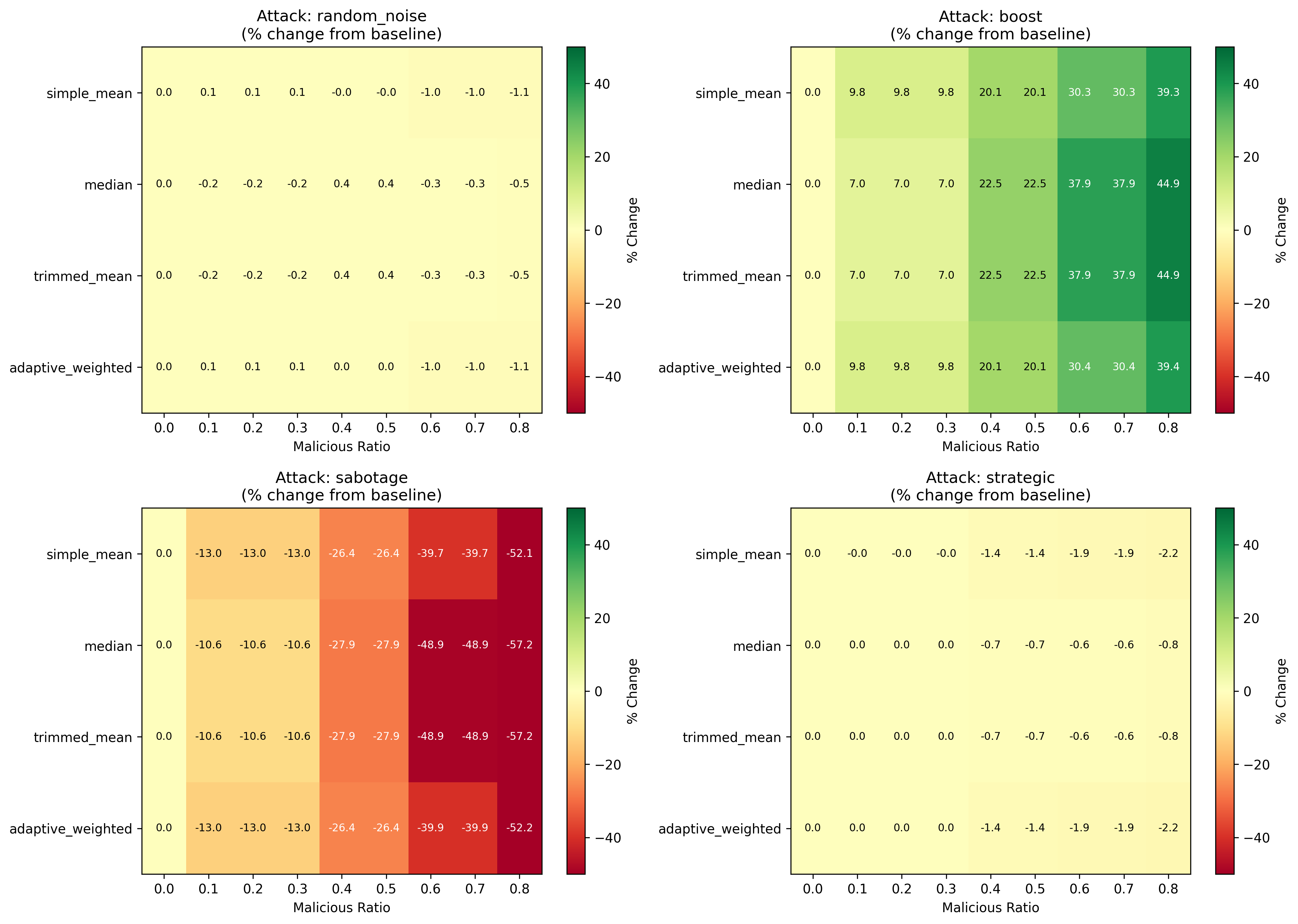}
\caption{Sensitivity heatmap over malicious ratio and defense method, aggregated across attack settings.}
\label{fig:sensitivity_ratio_heatmap}
\end{figure}

We evaluate robustness by injecting malicious evaluators following the threat model in Section~\ref{sec:system_threat}. Figure~\ref{fig:adversarial_defense} and Figure~\ref{fig:sensitivity_ratio_heatmap} visualize reward trends across attack types, malicious ratios, and defense rules. Table~\ref{tab:robustness_rho08} summarizes relative changes in average inference reward at malicious ratio $\rho = 0.8$, compared to the no attack setting for each defense rule.

\begin{table}[t]
\centering
\small
\setlength{\tabcolsep}{6pt}
\caption{Change in average inference reward at $\rho=0.8$ relative to $\rho=0.0$ for each attack type and defense rule. Positive change under boosting indicates payout inflation rather than improved quality.}
\label{tab:robustness_rho08}
\begin{tabular}{@{}l
S[table-format=+2.1]
S[table-format=+2.1]
S[table-format=+2.1]
S[table-format=+2.1]@{}}
\toprule
Attack & {Mean (\%)} & {Median (\%)} & {Trimmed (\%)} & {Adaptive (\%)} \\
\midrule
random\_noise & -1.1 & -0.5 & -0.5 & -1.1 \\
strategic     & -2.2 & -0.8 & -0.8 & -2.2 \\
sabotage      & -52.1 & -57.2 & -57.2 & -52.2 \\
boost         & +39.3 & +44.9 & +44.9 & +39.4 \\
\bottomrule
\end{tabular}
\end{table}

Robust aggregation mitigates noise style and strategic manipulations, where median and trimmed mean reduce reward shifts relative to mean under high malicious ratios. Sabotage attacks cause the largest degradation in inference reward across all defenses, indicating that strongly coordinated negative scoring remains a challenging regime for score based verification. Boosting increases inference rewards substantially, which is better interpreted as payout inflation driven by manipulated scores rather than improved output quality.

\subsection{Sensitivity to Evaluator Count $K$}
\label{subsec:results_sensitivity_k}

\begin{figure}[t]
\centering
\includegraphics[width=0.8\columnwidth]{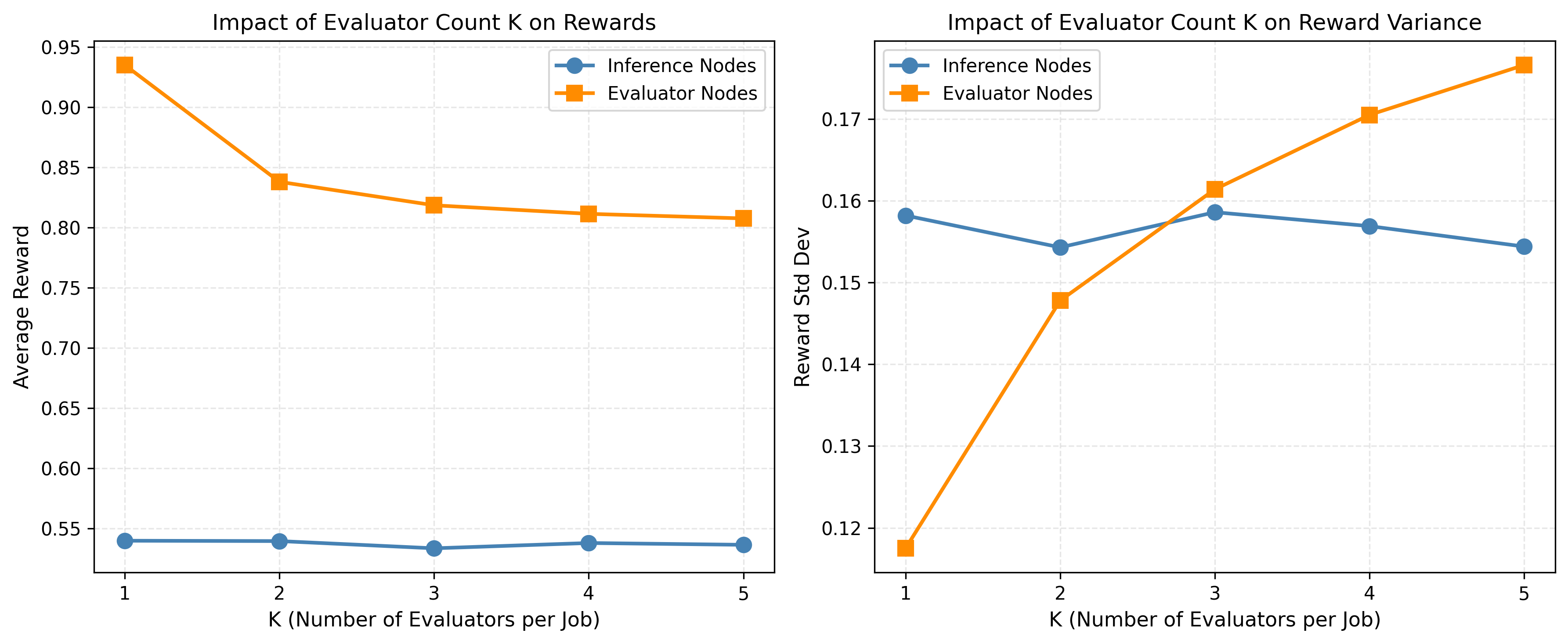}
\caption{Sensitivity of rewards to evaluator sample size $K$.}
\label{fig:sensitivity_k}
\end{figure}

We vary evaluator sample size $K$ to study the trade-off between robustness and evaluation overhead. Figure~\ref{fig:sensitivity_k} summarizes reward trends and variance across $K$. Table~\ref{tab:sensitivity_k} reports average inference reward, inference reward standard deviation, average evaluator reward, and evaluator reward standard deviation.

\begin{table}[htbp]
\centering
\small
\setlength{\tabcolsep}{6pt}
\caption{Sensitivity results over evaluator sample size $K$. Larger $K$ reduces evaluator average reward and increases evaluator reward variance, while inference reward remains relatively stable in this configuration.}
\label{tab:sensitivity_k}
\begin{tabular}{@{}c
S[table-format=1.3]
S[table-format=1.3]
S[table-format=1.3]
S[table-format=1.3]@{}}
\toprule
{$K$} & {Inf avg} & {Inf std} & {Eval avg} & {Eval std} \\
\midrule
1 & 0.540 & 0.158 & 0.935 & 0.118 \\
2 & 0.539 & 0.154 & 0.838 & 0.148 \\
3 & 0.534 & 0.159 & 0.818 & 0.161 \\
4 & 0.538 & 0.157 & 0.811 & 0.171 \\
5 & 0.536 & 0.154 & 0.808 & 0.177 \\
\bottomrule
\end{tabular}
\end{table}

Across the tested range, inference rewards remain stable, while evaluator rewards decrease as $K$ increases. This occurs because larger $K$ increases the chance that an evaluator deviates from consensus, which reduces closeness and therefore evaluator reward, and it also increases the dispersion of evaluator payoffs.

\subsection{Additional Plots}
\label{subsec:results_additional}

We provide additional visualizations that complement the main results, including defense summaries, weight evolution, scatter plots, and detailed malicious ratio comparisons. These plots can be included in the appendix or supplementary material.

%
%

\section{Discussion}
\label{sec:discussion}

This section interprets the empirical findings and highlights practical deployment implications for decentralized PoQ-style verification. We focus on robustness under adversarial evaluators, the role of aggregation and trust weighting, and the operational trade-offs introduced by evaluator sampling.

\subsection{When Robust Aggregation Outperforms Adaptive Weighting}
\label{subsec:discussion_robust_vs_adaptive}

A consistent pattern across our experiments is that robust aggregation rules provide immediate resilience against outliers and manipulation, even without additional state. Median and trimmed mean improve alignment with the ground truth proxy relative to simple mean, and they reduce reward sensitivity under noise and intermittent manipulations. This behavior is expected from robust statistics and Byzantine-resilient aggregation, where median-like rules limit the influence of extreme values \cite{huber1964robust,pmlr-v80-yin18a}.

In contrast, adaptive weighting is only effective when the trust update meaningfully separates reliable evaluators from unreliable ones. In our current implementation, trust weights are updated using deviation from the same consensus that they help define. This feedback loop can reduce the contrast between honest and malicious evaluators when attacks are symmetric or when malicious evaluators remain close to the consensus by design. In addition, small deviations across most rounds can cause many evaluator weights to drift upward until they saturate at the maximum trust bound, which reduces the effective difference between weighted and unweighted mean. In practice, this suggests two design directions. First, trust updates should be calibrated so that typical deviations do not lead to monotonic weight growth. Second, trust should ideally incorporate additional signals beyond consensus deviation, such as task-aware calibration, occasional anchor tasks with stronger supervision, or disagreement patterns across evaluator families.

\subsection{Payout Inflation and Distortion under Boosting Attacks}
\label{subsec:discussion_payout_inflation}

Boosting attacks increase submitted scores rather than decreasing them. Under such attacks, inference rewards can increase substantially, which is better interpreted as payout inflation rather than improved output quality. In a decentralized network, payout inflation is harmful because it transfers value to inference nodes without delivering commensurate quality, and it can also incentivize collusion between inference nodes and malicious evaluators.

This observation motivates reporting not only reward changes but also consensus alignment with an external quality proxy. When a defense rule produces higher rewards under attack, the correct interpretation depends on whether the consensus score remains aligned with the ground truth proxy. Robust aggregation can reduce the influence of boosted scores if the majority remains honest. However, at high malicious ratios, even robust rules can be overwhelmed because the median and trimmed mean depend on having a sufficiently large honest fraction. This is consistent with classical Byzantine assumptions that require an honest majority for certain aggregation guarantees \cite{lamport1982byzantine,castro1999practical}.

\subsection{Choosing Evaluator Sample Size $K$}
\label{subsec:discussion_k}

Evaluator sampling introduces a central operational trade-off. Larger $K$ increases the amount of evaluation information per job and can improve robustness in settings where outlier scores are common. However, larger $K$ also increases evaluation overhead and changes incentive dynamics for evaluator nodes. In our experiments, inference rewards remain relatively stable as $K$ varies, while evaluator rewards decrease and evaluator reward variance increases for larger $K$. This occurs because larger $K$ increases the chance that any given evaluator deviates from the consensus, which reduces closeness-based rewards.

From a deployment perspective, this suggests that $K$ should be selected based on a cost and risk budget. If evaluation is cheap and adversarial risk is high, choosing a larger $K$ can provide more robust consensus signals. If evaluation is expensive or evaluator participation needs stronger incentives, a moderate $K$ can reduce overhead and stabilize evaluator rewards. A practical approach is to set $K$ as a function of the observed malicious ratio estimate, increasing $K$ only when the system detects signs of instability or manipulation.

\subsection{Limitations and Future Work}
\label{subsec:discussion_limitations}

Our study has several limitations that motivate future work. First, the ground truth proxy uses a uniform token-level F1 score across tasks for offline analysis, which is appropriate for question answering but can underrepresent summarization quality dimensions. Future experiments should incorporate task-specific metrics such as ROUGE or semantic factuality measures. Second, the adversary model focuses on manipulation of evaluator scores and does not include more complex strategies such as identity sybils beyond a fixed malicious ratio. Sybil settings are a known challenge in decentralized systems and can undermine defenses that assume independent participants \cite{fung2020limitations}. Third, our trust update mechanism is a lightweight heuristic and can saturate under typical deviations. Future designs should explore more principled reliability estimation and calibration methods that remain robust under adaptive adversaries \cite{dawid1979em,raykar2010learningfromcrowds}.

Despite these limitations, our results show that robust aggregation provides a strong baseline defense and that evaluator heterogeneity must be treated as a first-class concern in PoQ-based verification mechanisms.

\section{Conclusion}
\label{sec:conclusion}

This extended paper studied adversary resilience in cost-aware Proof of Quality for decentralized LLM inference networks. We showed that evaluator heterogeneity is a central risk factor for score-based verification, and that naive averaging can degrade consensus alignment and reward stability when unreliable or malicious evaluators participate. We introduced robust aggregation rules and an adaptive trust weighting extension that preserve explicit cost awareness while improving tolerance to score manipulation.

Across correlation analysis and adversarial simulations, robust aggregation methods, including median and trimmed mean, consistently improved alignment with the ground truth proxy relative to simple mean and reduced sensitivity to noise and intermittent attacks. We also analyzed the operational role of evaluator sampling size $K$ and found a clear trade-off between evaluation overhead, evaluator incentives, and reward variance. These findings suggest practical guidelines for deploying PoQ-style verification, where robust consensus should be treated as a default component and where $K$ should be tuned to the risk and cost budget of the network.

Future work should incorporate task-specific quality proxies for summarization, strengthen trust updates with more principled reliability estimation, and evaluate sybil-resistant designs that remain robust when attackers can scale identities.

\bibliographystyle{plain}

\clearpage
\bibliography{references}

\appendix

\section{Appendix}
\label{sec:appendix}

\subsection{Additional Result Figures}
\label{subsec:appendix_additional_figs}

This appendix includes supplementary plots that provide deeper views into defense behavior, malicious ratio sensitivity, and trust weight dynamics. These figures complement the main results in Section~\ref{sec:results}.

\begin{figure}[htbp]
\centering
\includegraphics[width=\columnwidth]{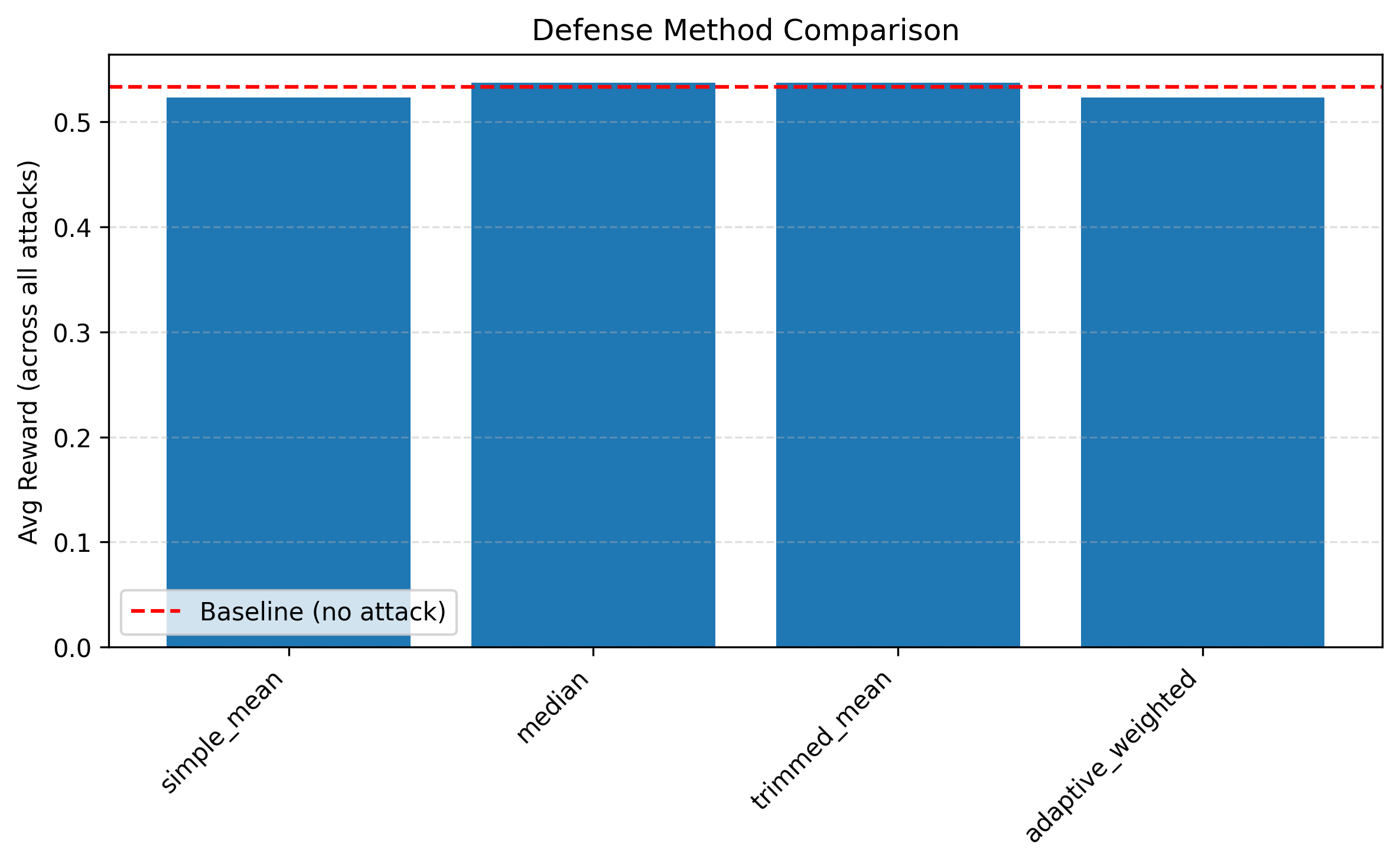}
\caption{Summary of defense performance aggregated across attack settings.}
\label{fig:defense_summary}
\end{figure}

\begin{figure}[htbp]
\centering
\includegraphics[width=\columnwidth]{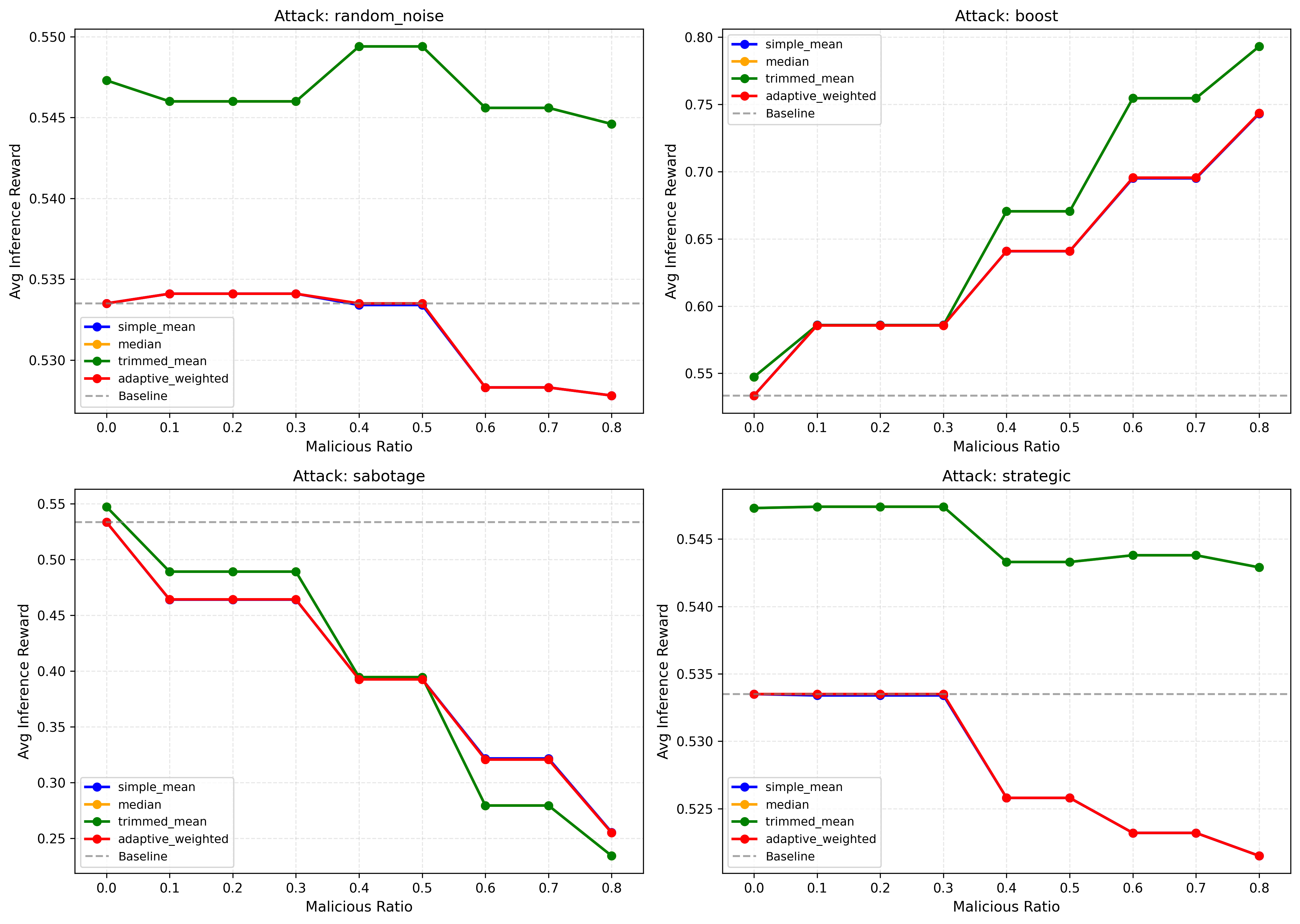}
\caption{Sensitivity to malicious ratio by attack type, comparing defense rules.}
\label{fig:sensitivity_ratio_by_attack}
\end{figure}

\begin{figure}[htbp]
\centering
\includegraphics[width=\columnwidth]{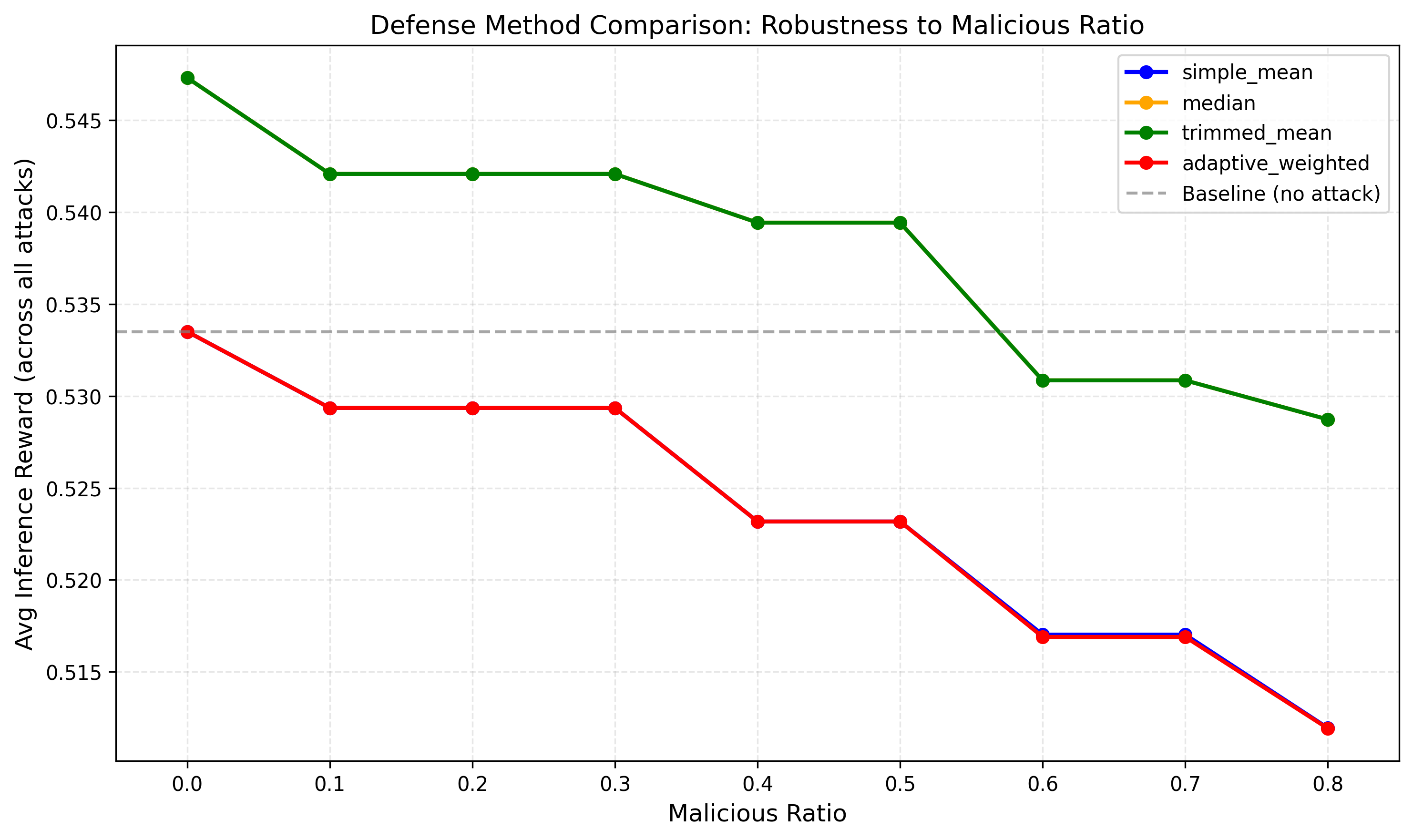}
\caption{Defense comparison as the malicious ratio increases. This figure highlights stability regimes and failure thresholds for each defense rule.}
\label{fig:sensitivity_ratio_defense_comparison}
\end{figure}

\begin{figure}[htbp]
\centering
\includegraphics[width=\columnwidth]{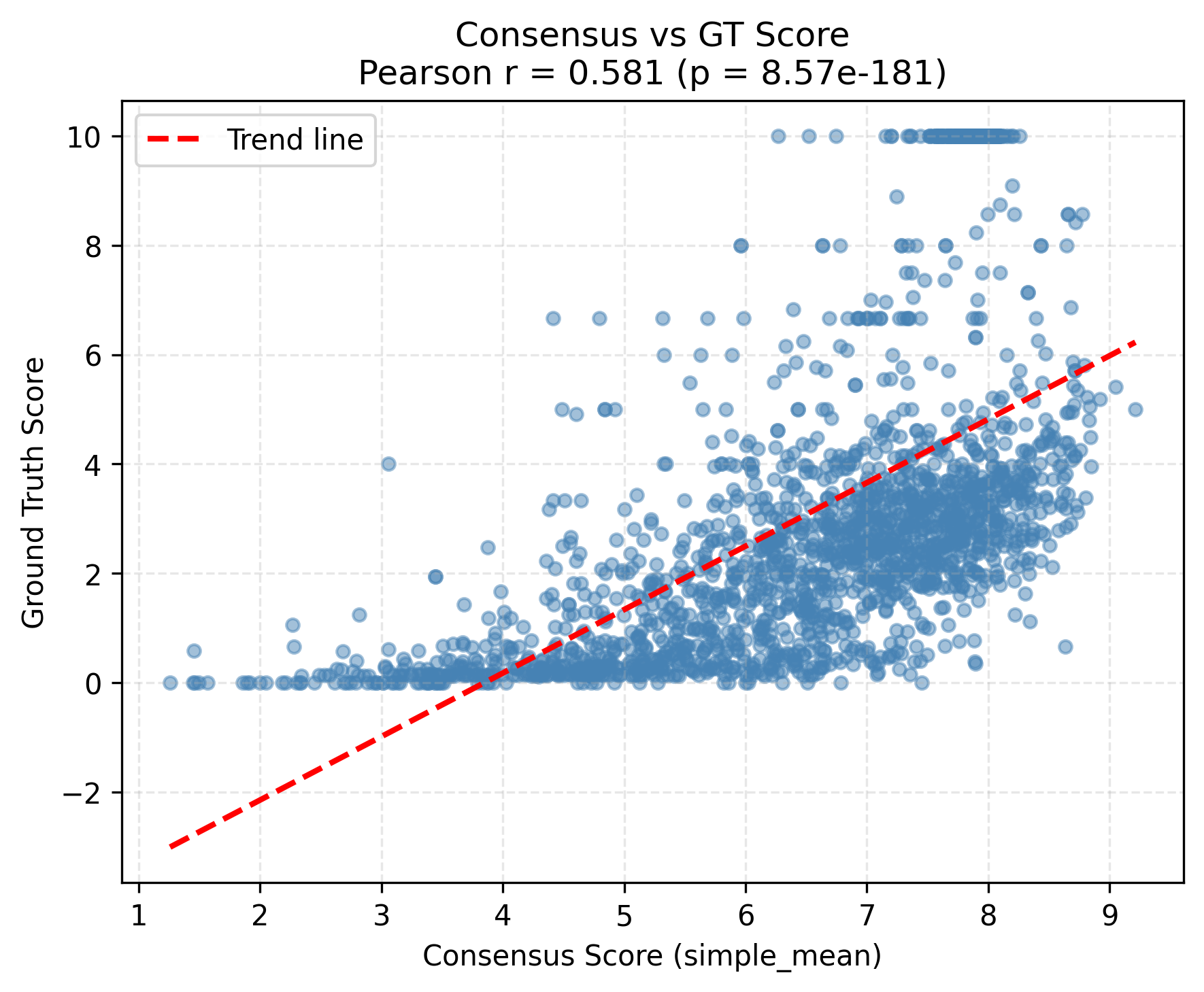}
\caption{Scatter plot view of evaluator scores and consensus scores against the ground truth proxy.}
\label{fig:correlation_scatter}
\end{figure}

\begin{figure}[htbp]
\centering
\includegraphics[width=\columnwidth]{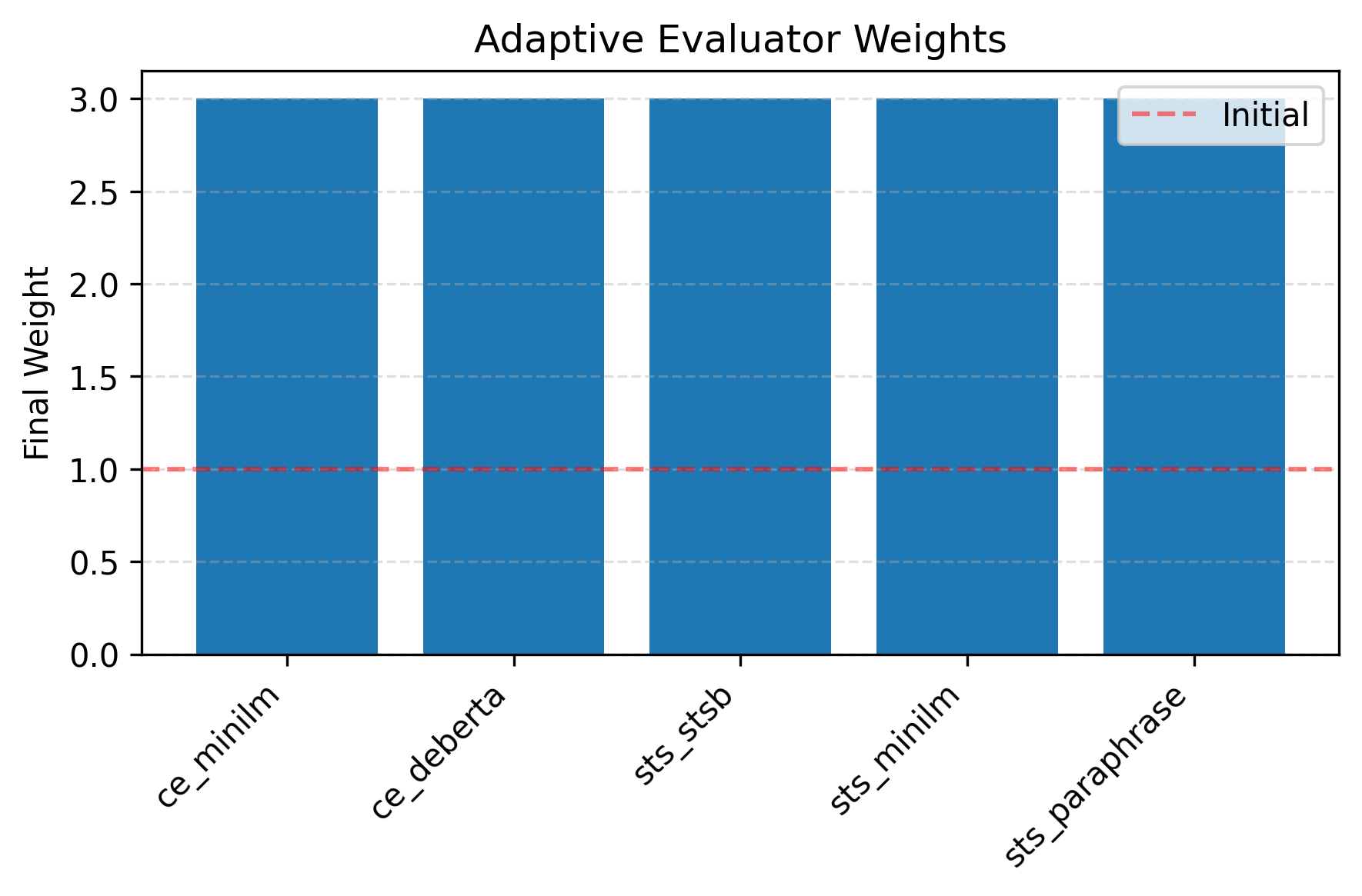}
\caption{Evolution of evaluator trust weights over simulation rounds under the adaptive weighted consensus rule.}
\label{fig:adaptive_weights}
\end{figure}

\subsection{Threat Model Parameters}
\label{subsec:appendix_threat_params}

For completeness, we list the adversarial parameters used in our experiments. Random noise injection uses a bounded uniform perturbation with range $[-r, r]$. Boosting and sabotage use a fixed bias $b$. Strategic manipulation applies a deviation magnitude $\Delta$ with probability $p$, and the deviation direction is sampled uniformly from $\{-1,+1\}$.

\subsection{Reproducibility Notes}
\label{subsec:appendix_reproducibility}

The full pipeline constructs tasks, generates model outputs, computes ground truth proxy scores, runs all evaluators, profiles latency for cost normalization, and then runs Monte Carlo simulations for baseline and adversarial settings. The simulation engine is modular with interchangeable consensus rules and attack models. All reported results are computed from stored JSON and CSV artifacts generated by the pipeline scripts.

\begin{table}[htbp]
\centering
\caption{Default protocol hyperparameters used in experiments.}
\label{tab:default_hparams}
\begin{tabularx}{\columnwidth}{lZ}
\hline
Hyperparameter & Value \\
\hline
$T$ (rounds) & 5000 \\
$K$ (evaluators per job) & 3 \\
Trim ratio $\gamma$ & 0.2 \\
Learning rate $\lambda$ & 0.1 \\
$w_{\min}, w_{\max}$ & 0.1, 3.0 \\
\hline
\end{tabularx}
\end{table}

\end{document}